# Spatio-Temporal Nonlinear Theory in Birefringent Microrings

Zelin Tan,[1, 2, †] Xingqiao Chen[1, 2], Ning Liu[1, 2], Jipeng Xu[1, 2], Jianfa Zhang[1, 2], Zhihong Zhu[1, 2], Ken Liu[1, 2, ‡] and Shiqiao Qin[1, 2, §]

[1]*College of Advanced Interdisciplinary Studies & Hunan Provincial Key Laboratory of Novel Nano Optoelectronic Information Materials and Devices, National University of Defense Technology, Changsha 410073, China.*

[2]*Nanhu Laser Laboratory, National University of Defense Technology, Changsha 410073, China.*

**Abstract:** Frequency-dependent nonlinear process in microresonators is widely acknowledged, but there is no theory available to calculate the conversion efficiency for each resonance of the ring, except for the phase-matching one. Similarly for azimuth-dependent nonlinear process in birefringent rings, there is a lack of theory to analysis the conversion efficiency for each azimuth of the ring. Consequently, it leads to old-fashioned or ill-considered coupling position and inefficient energy conversion in birefringent microrings. Here, we introduce spatio-temporal coupled-mode equation to describe mode spatial properties in the cavity, compensating for the deficiency of temporal coupled-mode equation in describing sophisticated responses. By this equation, we find that over a wide frequency range, the extremely strong second-harmonic generation can be achieved at different azimuth under different pumps in an X-cut lithium niobate microring, which is important for realizing an efficient entangled quantum light source, for example. This work will provide new ideas and guidelines for design and applications of monolithic birefringent photonic integrated circuits with high efficiency.

The study of nonlinear effects has a long history, since the invention of the laser in 1960s [1], evolving from bulk materials to integrated platform [2]. Beyond straight waveguides, no other device has received as much attention as microresonators in photonics integrated circuits, for their low footprint, high integration density and low power consumption [3]. Temporal coupled-mode equation (TCME) is commonly used to describe the nonlinear process in microrings, such as conversion efficiency [4,5], which works well in isotropic and phase-matched cases. However, in birefringent materials or other convoluted situation, the equation fails to characterize the sophisticated response, and it can only be explained simply by the spatial coupled-mode equation (SCME) in a compromise, which is used in straight waveguide regime [6-8]. For example, it is easy to understand that type-1 perfect phase matching can be achieved at four azimuths within a birefringent microring [9]. However, to the best of our knowledge, there is no theory to describe the nonlinear process in phase-mismatched and birefringent microresonators, even though the phenomenology is generally recognized. Here, we introduce the spatio-temporal CME compensating for the deficiency of TCME in describing sophisticated nonlinear responses within microresonators, such as conversion efficiency at different frequency pumps and azimuths.

In this work, we studied the second-harmonic generation (SHG) in birefringent microrings, one of the most complicated cases, as the refractive index and second-order nonlinear coefficient vary with angle. Both of these variations, one directly related to phase matching and one that will provide additional wavevectors due to the reversal of the polarization direction, can impose a significant impact on the nonlinear processes [7]. Thanks to the advancements made in the preparation and micro/nano-fabrication techniques of thin-film optical materials over the last few decades, materials with high nonlinear susceptibility, such as aluminum nitride [9-12], gallium nitride [13-15], gallium arsenide [16-18], and lithium niobate on insulator (LNOI) [19-21], have been extensively studied and developed for their potential applications in nonlinear optics. Among all the nonlinear materials, LNOI has been one of the most versatile and attractive platforms with exceptional nonlinear-, electro-, and acousto-optic properties, as well as its relatively high refractive index and wide transparency window [22]. Recent studies on periodically poling lithium niobate (PPLN) microresonators has prompted the conversion efficiency to an unprecedented level [23, 24]. However, most of these PPLN microrings are fabricated in Z-cut LNOI, in which the largest nonlinear tensor element $d_{33}$ (≈-27 pm/V) can only be leveraged under TM polarization, as well as the largest Pockels coefficient $r_{33}$ (≈31 pm/V). So, it introduces more obstacles during the fabrication of photonic integrated circuits (PIC). Although Z-cut EO modulators can be implemented by placing electrodes on top of the waveguides, it is difficult to efficiently deliver the electric field to sub-micrometer waveguides while maintaining low microwave and optical loss [22]. This is why X-cut LN platform is better suited for making complex PIC. An alternative approach is adapting racetrack-shape microcavities in X-cut LNOI, with periodically poled straight region [24]. This method is concise and efficient, but the influence of angle on conversion efficiency is neglected and fail to unleash the full potential of the device.

In X-cut LN microresonators, since the optical axis lies in the device plane, the nonlinear coefficient ($d_{33}$) and mode effective refractive index ($n_{\text{eff}}$) vary around the central axis due to birefringence. The inverted-sign $d_{\text{eff}}$ can introduce an additional wavevector K, similar to the periodic polarization inversion in a PPLN crystal [7]. Thus, it can compensate the phase mismatching between a certain high-order SHG mode and the fundamental pump mode. On the other hand, the variation of effective refractive index indicates that the conversion efficiency is dependent of azimuth besides frequency. In general, the physics of resonant optical systems can be described by the TCME, which is an analogy with a simple LC circuit [4], and derived by intuitive physics approach and the variational principle [5]. However, the spatial and momentum-related features of the mode are not included in TCMT, limiting the study of the cavity modes properties, like dispersion. In this letter, we perform a Taylor expansion of the angular frequency $\omega(k)$ at $\omega(k_0)$, where $\omega(k_0) = ck_0$, to obtain an equation for the angular frequency versus wavevector under dispersion regime. Then, it is substituted into TCME to get the final spatio-temporal CME.

With this equation, we can calculate the SHG conversion efficiency as a function of frequency (dispersion) and azimuthal angles, and it is shown that broadband, extremely strong SHG can be achieved in X-cut microrings at different angles. In the experiment, we measure the second-harmonic signal of two intracavity modes, with a 10-fold difference in conversion efficiency, which is consistent with the prediction of the equation. On the other hand, a CCD is used to capture the top-view optical microscopy image of the nonlinear signal scattering. There are 4 expected symmetrical bright scattering spots corresponding to 4 perfect phase-matching points. Distinct scattering positions of different pumps and the changing trend are also well in line with the theory.

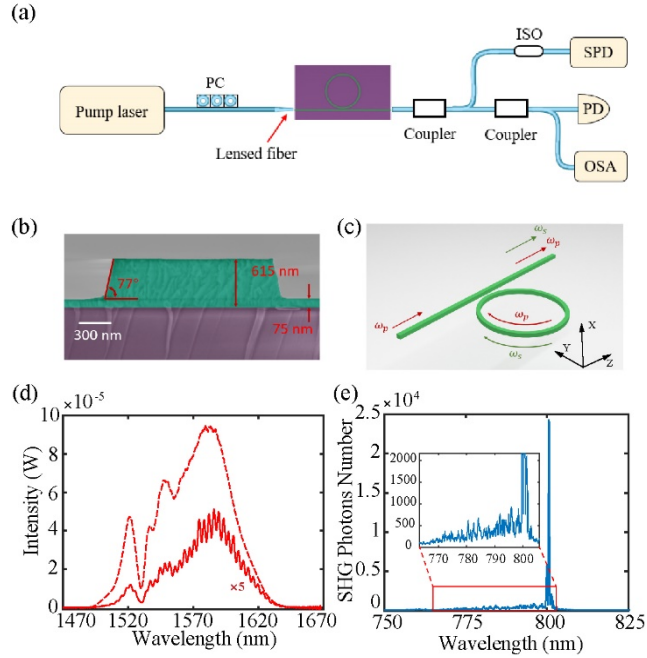

FIG. 1. (a) Illustration of the experimental setup of nonlinear optical processes in an on-chip microring. The pump, fs or continuous-wave (CW) laser, is coupled to the device by a lensed fiber. PC, polarization controller; ISO, isolator; SPD, single photon detector; PD, photodetector; OSA, optical spectrum analyzer. The transmission and second-harmonic spectra are measured by an OSA and a single-photon detector respectively. (b) False-color SEM image of the waveguide. (c) Schematic of doubly resonate SHG in an X-cut LN microring. $\omega_p$, $\omega_s$ are the frequencies of the pump and second-harmonic signal respectively. The coordinate system is same as that of the crystallographic. (d) Initial (dashed line) and transmission spectra (solid line) of fs laser. For ease of presentation, the transmission spectrum is multiplied by 5. (e) Second-harmonic spectrum excited by the fs laser in (d).

The experiment setup for characterizing the performance of microresonators and investigating the SHG process was schematically depicted in Fig. 1(a). A tunable continuous-wave diode laser (CW) or *fs* laser is used as the pump source, coupled into bus waveguide by a lensed fiber. One fiber PC was utilized to ensure the excitation of TE mode in the microring. When the pump is coupled into the microresonator, it will stimulate the doubly resonant second-harmonic generation. The second-harmonic signal is monitored by a single photon detector. A LN microring with radius of 21.9 μm is fabricated on a X-cut undoped LNOI wafer (~700 nm) by Argon plasma and bottom width of the ring waveguide is 2.2 μm. From the SEM in Fig. 1(b), we know that about 615 nm lithium niobate is etched, leaving a 75 nm slab. The waveguides have a slope angle of 77°. Fig. 1(c) illustrates the principle of SHG in an microcavity.

The coordinate axes show the standard crystallographic axes of LN, where the optical axis (z-axis) is in the plane and parallel to the direction of the straight waveguide. $\omega_p$, $\omega_s$ are the frequencies of the pump and second-harmonic signal, respectively, which satisfy the condition of energy conservation $2\omega_p = \omega_s$.

As mentioned before, broadband SHG can be achieved in birefringent microrings at the cost of reduced conversion efficiency, verified by a *fs* laser coupled into the ring in our experiment. The initial spectrum (dashed line) of the *fs* laser is shown in Fig. 1(d). From the transmission spectrum (solid line) in Fig. 1(d), it can be seen that a series of wavelengths are coupled into the microring, and they have a fixed spacing, called Free Spectral Range (FSR), which is roughly 1 THz. Also, it can be extrapolated from the intensities that the fiber-to-fiber coupling efficiency is about 10%. Fig. 1(e) shows the spectrum of the second-harmonic signal, measured by a single-photon detector. Obviously, the mode near 801 nm has a very pronounced SHG enhancement. The modes farther away have very weak conversion enhancement. Because the frequency of a *fs* laser is fixed, so the enhancement can be not as high as CW pump for detuning.

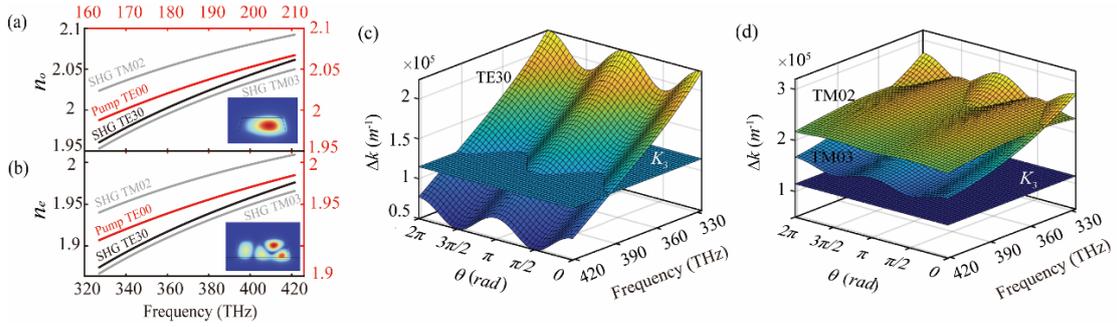

Fig. 2. Effective refractive indices of fundamental pump mode (red lines) and second-harmonic high-order modes (gray and black lines), $n_o$ (a) and $n_e$ (b), corresponding to the ordinary and extraordinary value at different frequency, simulated by COMSOL. The insets show the mode distribution of a fundamental pump mode (1601.8 nm) in (a) and its stimulated high-order second-harmonic mode (TE30) in (b). (c, d) Curved surfaces display the phase mismatching $\Delta k$ between the fundamental mode of pump and high-order second-harmonic modes, TE30 (c), TM02 (d) and TM03 (d), as a function of wavelength and angle. The plane is the third-order Fourier component of $d_{\text{eff}}$.

Although the results in Fig. 1(e) corroborate the previous conclusion, this is based on a fixed coupling position, neglecting the influence of azimuth. Next, we will derive the spatio-temporal CME to reveal the sophisticated nonlinear process in birefringent cavities from TCME. Suppose there are two modes in the LN microring in rotating frame, $E_i(t) = A_i(t)e^{-i\omega_i t}$, where $i$=1, 2 represent the fundamental mode of pump and one high-order mode of second-harmonic field. $A_i(t)$ is the normalized amplitude, such that $|A_i|^2$ is the number of photons of this mode in the cavity. So, the TCME for these two modes can be written as [6],

$$\frac{da_1}{dt} = (i\delta_1 - \frac{\kappa_1}{2})a_1 + 2i\gamma^* a_1^* a_2 e^{-i(\omega_2 - 2\omega_1)t} + \sqrt{\kappa_{1e}} s_{1+}, \qquad (1)$$

$$\frac{da_2}{dt} = -\frac{\kappa_2}{2}a_2 + i\gamma a_1^2 e^{i(-2\delta_1+\omega_2-2\omega_1)t}, \tag{2}$$

where $\delta_1 = \omega_p - \omega_1$, denotes the detuning of pump to the intracavity mode. $\gamma$ is the nonlinear coupling coefficient of two modes. $s_{i+}$ is the intensity of the driving source, and $\kappa_i = \kappa_{i0} + \kappa_{ie}$ is the total decay rate, in which $\kappa_{i0}$ ($\kappa_{ie}$) represents the intrinsic (coupling) decay rate of modes in the cavity. Both $\kappa_{i0}$ and $\kappa_{ie}$ can be obtained by fitting the transmission spectra of the cavity. For Eq. (1) the crossover term can be ignored since the two modes are weakly coupled. Assuming that there is no detuning $\delta_1 = 0$ between the pump and the intracavity mode, rotating-frame transformation is used again, and the TCME can be obtained finally:

$$\frac{da_1}{dt} = -\frac{\kappa_1}{2}a_1 + \sqrt{\kappa_{1e}}s_{1+}, \tag{3}$$

$$\frac{da_2}{dt} = -i\left(\Delta\omega + \frac{\kappa_2}{2}\right)a_2 + i\gamma a_1^2. \tag{4}$$

In Eq. (4), $\Delta\omega = \omega_2 - 2\omega_1$ represents the energy and phase mismatch between pump and second-harmonic modes. In principle, since the energy is conserved, $\Delta\omega = 0$. In this case, the above two equations cannot reflect the spatial characteristics of the intracavity mode caused by dispersion. To be able to show the effect of dispersion, we perform a Taylor expansion of $\omega(k)$ at $\omega(k_0)$, as follows:

$$\omega(k) = \omega(k_0) + \omega'(k_0)(k-k_0) + \frac{1}{2}\omega''(k_0)(k-k_0)^2 + o(k-k_0)^2. \tag{5}$$

The coefficient of the linear term, $\omega'(k_0) = c$, is speed of light in vacuum. This term indicates the spatial characteristics of a traveling wave. For a standing wave with parabolic dispersion, it is characterized by the second-order term with coefficient $\omega''(k_0)$ [25]. The wavevector $k = \frac{2\pi n_{eff}}{\lambda}$, as $n_{eff}$ is the effective refractive index of the modes in a X-cut lithium niobate microring, which satisfied

$$n_{eff}(\theta) = \frac{n_o n_e}{(n_e^2 \sin^2\theta + n_o^2 \cos^2\theta)^{\frac{1}{2}}}. \tag{6}$$

$\theta$ is the angle between polarization and optical axis (z). $n_o, n_e$ denote the ordinary and extraordinary effective refractive index of LN, which can be simulated by COMSOL respectively, shown in Fig. 2(a, b), with an isotropic definition of material refractive index. The red line shows the value of the pump mode at different frequency, while black (TE30) and gray lines represent that of the second-order high-order modes.

The modes in microresonator can be described by the first-order approximation in Eq. (5) as a traveling wave. So, take the first two terms in the expansion, and we get

$$\Delta\omega(k) = \omega(k_2) - 2\omega(k_1) + c(k_{22} - k_2) - 2c(k_{11} - k_1). \tag{7}$$

In this equation, $k_1$ ($k_2$) is the wavevector of pump (second-harmonic) mode without dispersion, so $k_2 = 2k_1$, while $k_{11}$ ($k_{22}$) is the wavevector of the actual pump (second-harmonic) mode in dispersion system. Thus, Eq. (7) can be simplified to

$$\Delta\omega(k) = c(k_{22} - 2k_{11}) = \frac{2\pi c \Delta n}{\lambda_{22}} = c\Delta k, \tag{8}$$

where $\Delta n$ can be obtained by substituting the calculated refractive index $n_o, n_e$ in Fig. 3(a, b) into Eq. (6), then doing subtracting between the picked modes. Finally, substituting Eq. (8)

into Eq. (4), we get the spatio-temporal CME for SHG in X-cut LN microresonators without detuning,

In X-cut LN, the second-order nonlinear coefficient can be expressed as [7]:
$$d_{\text{eff}} = -d_{22}\cos^3\theta + 3d_{31}\cos^2\theta\sin\theta + d_{33}\sin^3\theta, \tag{9}$$
with the same definition of $\theta$ in Eq. (6). In a ring, $\theta = l/r$, where $l$ is the distance light propagates in the ring, and $r$ represents the effective radius of the ring. Since the sign of $d_{\text{eff}}$ inverts periodically, it will provide an additional wavevector $K_i$ ($i$ denotes $i$th-order Fourier component of the $d_{\text{eff}}$ in the wavevector space), like PPLN. This additional wavevector may compensate for the phase mismatch between the pump mode and a higher-order second-harmonic mode, resulting in perfect phase matching. It is clear that there are two Fourier components of $d_{\text{eff}}$, $K_1 = 1/r$ and $K_3 = 3/r$. Curved surfaces in Fig. 2 (c, d) show the phase mismatch $\Delta k$ versus wavelength and azimuth for different pump mode and higher-order modes of second-harmonic field, TE30 (c), TM03 (d), and TM02 (d). The plane in the figure represents the third-order Fourier component of $d_{\text{eff}}$, $K_3$. It is shown that only when the second-harmonic mode is TE30, can the phase mismatch between it and the TE fundamental mode of the pump be canceled out by $K_3$ provided by $d_{\text{eff}}$. The perfect phase matching is achieved at some angles, since $\Delta n$ is related to the $sin$ function. The wavelength range where this perfect-phase matching can be achieved in certain azimuth is roughly 1525 nm-1603 nm from Fig. 3(c).

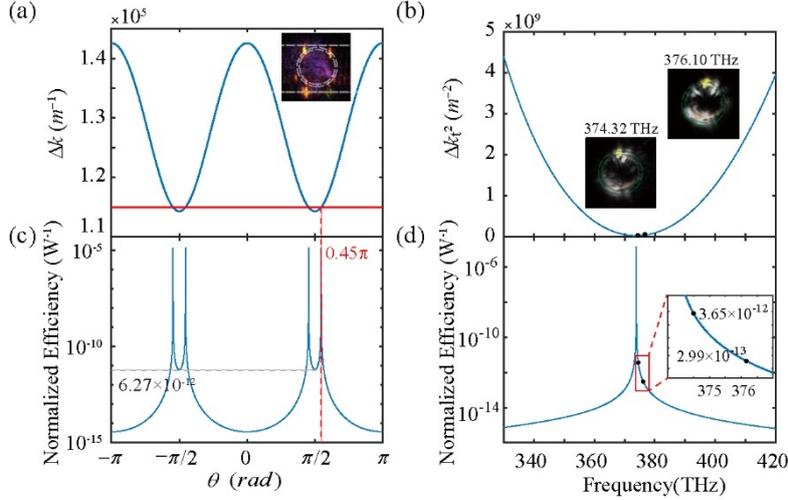

FIG. 3. Normalized conversion efficiency versus azimuth and frequency. (a) Blue line is the phase mismatch between fundamental mode of the pump (1601.8 nm) and its second-harmonic mode (TE30), while red line is $K_3$ provided by $d_{\text{eff}}$. Inset: Top-view optical microscopy image of the nonlinear signal scattering. (b) At $\theta = \pi/2$, the square of total phase mismatch at different frequency. Insets: Top-view optical microscopy images of the nonlinear signal scattering under different pump. (c, d) Normalized conversion efficiency calculated from (a) and (b).

The effect of azimuth and frequency on conversion efficiency can be derived from the spatio-temporal CME. In the non-depletion pump regime with weak coupling between modes, the steady state satisfies $da_1/dt = da_2/dt = 0$. Therefore, from Eqs. (5, 6), we can get the

normalized second-harmonic conversion efficiency without detuning:

$$\Gamma \equiv \frac{P_2}{P_1^2} = \frac{|\gamma|^2 \omega_2}{\hbar \omega_1^2} \frac{\kappa_{1e}^2}{(\kappa_1/2)^4} \frac{\kappa_{2e}}{(\kappa_2/2)^2 + (c\Delta k_t)^2}, \qquad (10)$$

where $\Delta k_t = \Delta k - K_i$, denoting the total phase mismatch. Fig. 3 (a) shows the variation of $\Delta k$ (blue line) with azimuth for pump of 1601.8 nm and the wavevector $K_3$ (red line) provided by $d_{\text{eff}}$. Substituting them into Eq. (10), the conversion efficiency as a function of angle by this pump can be obtained with constant decay rates, as shown in Fig. 3(c) (without considering the coupling coefficient $|\gamma|^2$). The maximum conversion efficiency occurs at $|\theta| = 0.45\pi$. In this experiment, the coupling point between bus waveguide and ring is at $\theta = \pi/2$, depicted in Fig. 1(c). Therefore, the optimal conversion is not achieved at this occasion. The four bright scattering points in the inset of Fig. 3(a) around the ring indicate four perfectly matched points within the ring. Fig. 3(b) displays the total phase mismatch $\Delta k_t^2$ with frequency at $\theta = \pi/2$. Perfect phase matching was achieved when the pump was around 1603 nm, (as the second-harmonic mode at about 374 THz). The corresponding conversion efficiency is shown in Fig. 3(d). Insets in Fig. 3(b) show two CCD photographs of the signal generated by the nonlinear process of different pumps in the ring. White shows near-infrared second-harmonic signal near 800 nm, and yellow shows third-harmonic signal near 500 nm. Because the perfect phase-matching point is different for each pump as the intersections in Fig. 3(a), the position of the strongest SHG and its scattering is also different, just as the white dots in two images are in different positions. For the SHG at 376.10 THz, the distance between the two perfect-match points (in the upper part of the ring) is greater than that at 374.32 THz, since the $\Delta k$ decreases with frequency shown in Fig. 2 (a, b), which is also in good agreement with theory. The blurred white area at the bottom of the image is due to scattering between the straight waveguide and the ring. Black points in Fig. 3(d) are simulated data for two intracavity modes at 1601.8 and 1954.2 nm. In the experiment, second-harmonic signal are measured of these two cavity modes as shown in Fig. 4(a) as well as their transmission spectra in Fig. 4(b). There's almost a tenfold difference in intensity, which is in agreement with the theory in Fig. 3(d).

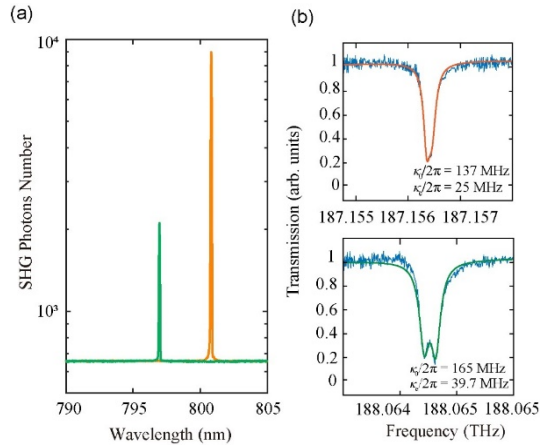

FIG. 4. (a) SHG spectra of two intracavity modes, 1601.8 nm (orange) and 1594.2 nm (green). (b) Transmission spectra of the two cavity modes in (a). Orange (green) line is the Lorentz fitting of the resonance for extracting decay rate.

To summarize, we propose the spatio-temporal CME for the inclusion of mode space properties (dispersion) in a microring by performing a Taylor expansion of the angular frequency term in TCME, for the first time to our best knowledge. With this equation we can calculate the conversion efficiency at different pumps and positions on the ring, especially which is significant for birefringent platform, like X-cut LN. Because in the X-cut LN or other birefringent ring, due to the additional wavevectors provided by the second-order nonlinear coefficients and azimuth-dependent effective index, a perfect phase matching will be achieved in a specific range of frequencies and at different azimuth, which will result in extremely strong SHG. So, this equation will play a brilliant role in the design of birefringent nonlinear devices with high efficiency, such as the settlement of coupling position between ring and bus waveguide. Constrained by fabrication period, we didn't perform the coupling at different azimuth in this paper to verify the theory directly, and took top-view CCD photograph at second hand instead. However, this work will help us to optimize the design and application of monolithic LN PIC for higher energy efficiency in the future, as well as the other well-developed integrated birefringent platform.

We thank Wei Liu and Jiacheng Liu for valuable discussions during the preparation of this manuscript. This project was supported by the National Natural Science Foundation of China (12274462, 11674396), Hunan Provincial Science and Technology Department, China (2017RS3039, 2018JJ1033), Hunan Provincial Innovation Foundation for Postgraduate, China (QL20210006) and Hunan provincial major sci-tech program (2023zk1010).

[†] zelin_tan@163.com

[‡] liukener@163.com

[§] sqqin8@nudt.edu.cn

[1] P. A. Franken, A. Hill, C. W. Peters, and G. Weinreich, *Generation of Optical Harmonics*, Physical Review Letters **7**, 118 (1961).

[2] P. Liu, H. Wen, L. Ren, L. Shi, and X. Zhang, *χ(2) Nonlinear Photonics in Integrated Microresonators*, Front. Optoelectron. **16**, 18 (2023).

[3] R. A. Soref, *The Past, Present, and Future of Silicon Photonics*, IEEE Journal of Selected Topics in Quantum Electronics **12**, 1678 (2006).

[4] H. A. Haus, *Waves and Fields in Optoelectronics* (1983).

[5] H. A. Haus and W.-P. Huang, *Coupled-Mode Theory*, Proceedings of the IEEE **79**, 1505 (1991).

[6] R. Luo, Y. He, H. Liang, M. Li, J. Ling, and Q. Lin, *Optical Parametric Generation in a Lithium Niobate Microring with Modal Phase Matching*, Physical Review Applied **11**, (2019).

[7] J. Lu, M. Li, C.-L. Zou, A. A. Sayem, and H. X. Tang, *Toward 1% Single-Photon Anharmonicity with Periodically Poled Lithium Niobate Microring Resonators*, Optica **7**, 1654 (2020).

[8] J. Lin et al., *Broadband Quasi-Phase-Matched Harmonic Generation in an On-Chip Monocrystalline Lithium Niobate Microdisk Resonator*, Physical Review Letters **122**, (2019).

[9] G. Lin, J. Fürst, D. Strekalov, and N. Yu, *Wide-Range Cyclic Phase Matching and Second Harmonic Generation in Whispering Gallery Resonators*, Applied Physics Letters **103**, (2013).

[10] J.-Q. Wang, Y.-H. Yang, M. Li, X. Hu, J. B. Surya, X.-B. Xu, C. Dong, G. Guo, H. X. Tang, and C. Zou, *Efficient Frequency Conversion in a Degenerate χ(2) Microresonator*, Physical Review Letters **126**, (2021).

[11] X. Guo, C.-L. Zou, and H. X. Tang, *Second-Harmonic Generation in Aluminum Nitride Microrings with 2500%/W Conversion Efficiency*, Optica **3**, 1126 (2016).

[12] M. Soltani, R. A. Soref, T. Palacios, and D. Englund, *AlGaN/AlN Integrated Photonics Platform for the Ultraviolet and Visible Spectral Range*, Optics Express **24**, 25415 (2016).

[13] A. Bruch, C. Xiong, B. Leung, M. Poot, J. Han, and H. X. Tang, *Broadband Nanophotonic Waveguides and Resonators Based on Epitaxial GaN Thin Films*, Applied Physics Letters **107**, (2015).

[14] M. S. Mohamed, A. Simbula, J. -f. Carlin, M. Minkov, D. Gerace, V. Savona, N. Grandjean, M. Galli, and R. Houdré, *Efficient Continuous-Wave Nonlinear Frequency Conversion in High-Q Gallium Nitride Photonic Crystal Cavities on Silicon*, APL Photonics **2**, (2017).

[15] Y.-Z. Zheng, C. Sun, B. Xiong, L. Wang, Z. Hao, J. Wang, Y. Han, H. Li, J. Yu, and Y. Luo, *Integrated Gallium Nitride Nonlinear Photonics*, Laser & Photonics Reviews **16**, 2100071 (2021).

[16] P. S. Kuo, J. Bravo-Abad, and G. S. Solomon, *Second-Harmonic Generation Using -Quasi-Phasematching in a GaAs Whispering-Gallery-Mode Microcavity*, Nature Communications **5**, (2014).

[17] P. S. Kuo, W. Fang, and G. S. Solomon, *4-Quasi-Phase-Matched Interactions in GaAs Microdisk Cavities*, Optics Letters **34**, 3580 (2009).

[18] C. P. Dietrich, A. Fiore, M. G. Thompson, M. Kamp, and S. Höfling, *GaAs Integrated Quantum Photonics: Towards Compact and Multi-Functional Quantum Photonic Integrated Circuits*, Laser & Photonics Reviews **10**, 870 (2016).

[19] J. Lu, A. A. Sayem, Z. Gong, J. B. Surya, C.-L. Zou, and H. X. Tang, *Ultralow-Threshold Thin-Film Lithium Niobate Optical Parametric Oscillator*, Optica **8**, 539 (2021).

[20] C. Wang, M. Zhang, X. Chen, M. Bertrand, A. Shams-Ansari, S. Chandrasekhar, P. J. Winzer, and M. Loncar, *Integrated Lithium Niobate Electro-Optic Modulators Operating at CMOS-Compatible Voltages*, Nature **562**, 101 (2018).

[21] M. He et al., *High-Performance Hybrid Silicon and Lithium Niobate Mach–Zehnder Modulators for 100 Gbit S−1 and Beyond*, Nature Photonics **13**, 359 (2019).

[22] D. Zhu et al., *Integrated Photonics on Thin-Film Lithium Niobate*, Advances in Optics and Photonics **13**, 242 (2021).

[23] J. Lu, J. B. Surya, X. Liu, A. Bruch, Z. Gong, Y. Xu, and H. X. Tang, *Periodically Poled Thin-Film Lithium Niobate Microring Resonators with a Second-Harmonic Generation Efficiency of 250,000%/W*, Optica **6**, 1455 (2019).

[24] J. Chen, Z. Ma, Y. M. Sua, Z. Li, C. Tang, and Y.-P. Huang, *Ultra-Efficient Frequency Conversion in Quasi-Phase-Matched Lithium Niobate Microrings*, Optica **6**, 1244 (2019).

[25] A. C. Overvig, S. A. Mann, and A. Alù, *Spatio-Temporal Coupled Mode Theory for Nonlocal Metasurfaces*, ArXiv (Cornell University) (2023).